# Automating the Communication of Cybersecurity Knowledge: Multi-Case Study

Alireza Shojaifar[1,2], Samuel A. Fricker[1,3], Martin Gwerder[1]

[1]FHNW, IIT and IMVS, 5210 Windisch, Switzerland
(alireza.shojaifar|samuel.fricker|martin.gwerder)@fhnw.ch
[2]Utrecht University, Dept. of Information and Computing Sciences, Utrecht, Netherlands
a.shojaifar@uu.nl
[3]Blekinge Institute of Technology, SERL-Sweden, 371 79 Karlskrona, Sweden
samuel.fricker@bth.se

**Abstract.** Cybersecurity is essential for the protection of companies against cyber threats. Traditionally, cybersecurity experts assess and improve a company's capabilities. However, many small and medium-sized businesses (SMBs) consider such services not to be affordable. We explore an alternative do-it-yourself (DIY) approach to bringing cybersecurity to SMBs. Our method and tool, CYSEC, implements the Self-Determination Theory (SDT) to guide and motivate SMBs to adopt good cybersecurity practices. CYSEC uses assessment questions and recommendations to communicate cybersecurity knowledge to the end-user SMBs and encourage self-motivated change. In this paper, the operationalisation of SDT in CYSEC is presented and the results of a multi-case study shown that offer insight into how SMBs adopted cybersecurity practices with CYSEC. Effective automated cybersecurity communication depended on the SMB's hands-on skills, tools adaptedness, and the users' willingness to documenting confidential information. The SMBs wanted to learn in simple, incremental steps, allowing them to understand what they do. An SMB's motivation to improve security depended on the fitness of assessment questions and recommendations with the SMB's business model and IT infrastructure. The results of this study indicate that automated counselling can help many SMBs in security adoption.

**Keywords:** Cybersecurity, Small and Medium-sized Businesses, Capability Assessment and Improvement, Do-It-Yourself, Multi-Case Study.

## 1 Introduction

Small and medium-sized businesses (SMBs) as the foundation of the EU's economy [1] are the weakest spot for cyber-attacks [2,3]. SMBs have specific characteristics, and these characteristics separate them from large companies and make them highly vulnerable to security attacks [4,5,6 ,7]. The lack of financial resources, expertise, written formal security policies, and also the common wrong attitude towards security and risks are some of these characteristics. Previous studies considered these characteristics and their influences on SMBs' resilience to security threats [3,8,9,10].

Ntouskas et al. [3] present a self-management security method which provides a consultancy environment for SMBs. Brunner et al. [9] focus on the level of automation in information security management and describe a continuous, risk-driven, and context-aware information security management system. Their framework is applicable to SMBs [9]. Furnell et al. [10] present a self-paced, flexible, and personalised security training software tool. The tool provides employees with the ability to learn some security countermeasures and desired behaviours.

In our experience of working with SMBs that were active in the IT industry, we found out that access to knowledge is not enough for motivating the SMBs to adopt appropriate behaviour. The SMBs need to understand the severity of threats and the impacts on their businesses. Moreover, providing hands-on skills that are consonant with the SMBs' capability motivates them to have security practices. We have not found any approach that considers the SMBs' motivation in the adoption of cybersecurity through self-assessment, learning, and improvement. Technical security measures form a large part of information security research [11]. Cybersecurity is more effective if the attention goes beyond the technical protecting means to the users, social, and organisational environment [11,12,13]. Human errors are the main cause of security failures [12] and promoting users' self-efficacy, and knowledge in information security can enhance organisation security [14].

CYSEC is a self-paced SMB-specific training and assessment method that automates elements of a counselling dialogue [15] between a security expert and employees in the SMB to ward off cyber threats. The interaction and dialogue between employees and security experts bridge the gap between them and makes the information security measures more effective [16]. Since users' resistance to accepting security tools and advice is one of the main problems for information security [17], the dialogue in CYSEC is based on theoretical foundations of motivation and the effects of employees' psychological needs on cybersecurity adoption. Persuasion is more effective than rational training strategies when the level of commitment to change is low [18].

The current study focuses on CYSEC evaluation to see whether the CYSEC is useful and effective as a method of communicating cybersecurity expertise for enabling DIY cybersecurity assessment and capability improvement for SMBs. The study purpose was approached by using an observation strategy based on the think-aloud protocol. While the previous literature mainly studies individuals' security behaviour through several interviews [16,19,20], the empirical findings of this study are based on observing actual usage of the tool to determine those factors which facilitate or control users' behaviour. The data was qualitatively analysed based on our theoretical model derived from Self-determination Theory (SDT) [21,22]. Our results demonstrate that SDT can explain motivational factors for effective counselling communication, and these factors influence users' behaviour to adopt cybersecurity recommendations. Unmet psychological needs may hamper users' adoption of cybersecurity behaviours. We observed that the automated dialogue is more effective when the method offers adapted behaviour, users' self-efficacy improvement, and SMBs' confidentiality issues together.

The remainder of the paper is structured as follows. Section 2 presents the theoretical background. Section 3 describes the CYSEC method. Section 4 describes the design of the study. Section 5 presents the process of data collection in SMBs. Section 6 analyses the results and answers the research questions. Section 7 discusses the significance of the results and the threats to validity. Section 8 summarises and concludes.

## 2    Theoretical Background

Cybersecurity studies draw on a variety of theories from different disciplines [19,20]. Self-Determination Theory (SDT) [21] provides a rigorous theoretical framework for studying motivation and has been considered in cybersecurity [19,20]. SDT describes and explains people's psychology of being self-motivated for adopting personal behaviours [21]. SDT was developed and evaluated with extensive research that resulted in an in-depth understanding of the conditions under which people will develop towards being a self-motivated in pursuing what they and their community consider as being desirable. The results of the research help managers and coaches to bring meaningful norms of behaviour into use and support the concerned people in adopting the conduct.

Self-motivation concerns goal-orientation, energy, and persistence – all related to producing results. If a goal is perceived to be important, the concerned person will start adapting his or her behaviour and be persistent to the extent that the behavioural change will sustain. According to SDT, a person will be self-motivated if these psychological needs have been satisfied: competence, autonomy, and relatedness. A lack of perceived competence, or self-efficacy, will lead the person to give up. Autonomy is important as the free choice determines how convinced the person is about the behaviour to be adopted. Relatedness to a person who acts as a role model for the behaviour can reinforce the self-motivation and even offer a template of how to adopt the behaviour.

Both intrinsic and extrinsic motivation leads to the adoption and internalisation of new behaviour. However, the more intrinsic the motivation is, the more effective and sustainable the adoption of the behaviour is. For each type of motivation, several forces influence how people are moved to act. People can feel motivated because they value an activity, e.g., by an abiding interest. People with such intrinsic motivation have interest, excitement, and confidence, which manifests as enhanced performance, persistence, and creativity. People under external coercion, e.g. with a bribe, fear of being surveilled, or other external pressure, are risking to be unwilling and unmotivated. Still, people can be externally motivated by a stimulating personal commitment to excel and offering role models' recognition. Table 1 shows, for the continuum from intrinsic motivation to amotivation, how behaviour may be influenced. Any method for helping users to achieve goals should operationalise these factors in the method's design.

Table 1 is pointing to the important SDT constructs that should be operationalised by a coaching method. It suggests hypotheses that can be used for evaluating whether the method supports the effectiveness of the cybersecurity knowledge communication for SMBs. The constructs concern attributes of the method user and environment with which the user interacts. The method user's attributes characterise the user's desired behaviour, self-efficacy, and autonomy. The method environment's attributes are relatedness, belonging, and connectedness offered to the user, pressure imposed through rewards, threats, and deadlines, the knowledge provided for helping the user to develop self-efficacy, and choice offered for fostering autonomy of the user.

**Table 1.** Factors for influencing desired behaviour, based on SDT.

| Motivation | How Desired Behaviour is Influenced |
|---|---|
| **Intrinsic motivation**: a person with interested and joy in desired behaviour tends to seek out novelty and challenges, to explore, learn, and exercise one's capacities even in the absence of specific rewards. | Autonomy of choice, perceived competence or self-efficacy, and a caring environment with optimal challenges and feedback of how the person's actions lead to the outcomes enhance intrinsic motivation and performance [23]. Extrinsic rewards, threats, deadlines, pressured evaluations, and imposed goals diminish intrinsic motivation. |
| **Extrinsic motivation**: continuum from coercion to stimulating intrinsic motivation: <br> **A) External regulation** is associated with control or alienation, and actions are perceived imposed by external regulators <br> **B) Introjected regulation** is not accepted as the one's own, but behaviours are performed to maintain a feeling of worth, e.g. to avoid guilt or anxiety or attain pride <br> **C) Regulation through identification**: conscious valuing and acceptance of rules as being personally important <br> **D) Integrated regulations:** fully assimilated as a result of evaluation and brought into congruence with one values and needs | With prescribed behaviours and values, new behaviour is internalised with meaningful rationales, autonomy, and relatedness [23]. External regulation is achieved with salient rewards or threats. Introjected regulation is achieved with the provision of belonging and connectedness, e.g. by having significant others to whom people feel attached or related prompt, model, endorse, or value the desired behaviour. Regulation through identification can only be achieved if autonomy of choice is provided. To integrate a regulation, the rules' meaning must be synthesised with respect to the person's goals and values with great autonomy in the sense of choice, volition, and freedom from excessive external pressure. |
| **Amotivation**: lacking the intention to act due to coercion, leading to failed goal achievement. | Amotivation is resulting from not valuing an activity, not feeling competent to do it, or not expecting the activity to yield desired outcome. |

## 3  CYSEC, a DIY Cybersecurity Improvement Method

CYSEC is a method and tool allowing SMBs' Chief Information Security Officer (CISO) to improve cybersecurity in a do-it-yourself fashion. The method guides the CISO in following Deming's plan-do-check-act (PDCA) [24] cycles of selecting sensible security themes, implementing a recommended practice, checking progress, and adapting based on lessons-learned. The tool offers memory allowing the CISO to continue the PDCA work where he left off. The tool also includes design elements based on SDT that aim at offering motivation for effective results and sustainability of the improvements.

Fig. 1. shows the two main interfaces offered to the user. A dashboard offers the features (1) recommendations for next improvements, (2) access to capability areas for PDCA work, (3) summary information about the company progress. Once the PDCA work for a given capability area is started, e.g. by choosing a recommendation or a capability area, the user enters the work area that offers the features (4) self-assessment, (5) access to expert knowledge, and (6) action cockpit for creating calendar entries, emails, and reminders. Table 2. describes how CYSEC operationalises SDT.

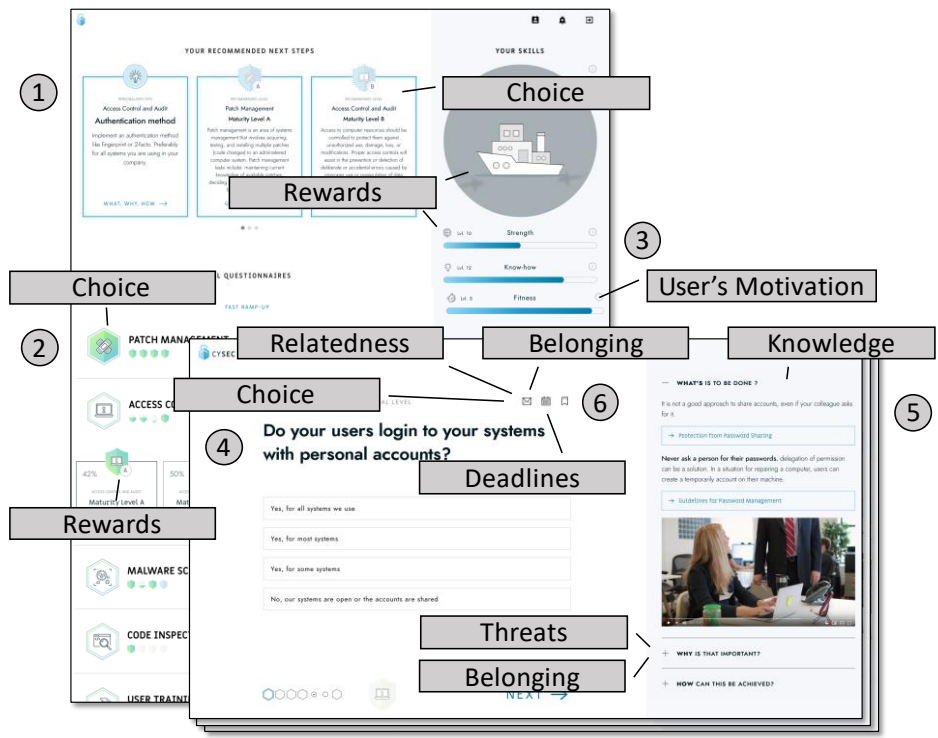

**Fig. 1.** Main user interfaces of the CYSEC tool and mapping of its features to SDT constructs.

The content has been organised into five cybersecurity themes (Patch Management, Access Control and Audit, Malware Scans, User Training, Back up). These themes allow fast ramp-up of security capabilities with minimal effort and large impact on SMBs.

Recommendations are generated based on users' answers to the self-assessment questionnaires for maturity improvement [25]. For the first time, new users will see one recommendation to fill out the company coach. As an adaptation rule, the answers to the company coach affect the questions asked in the other coaches. After completing the company coach, several coaches will be active in the dashboard. The available capabilities are defined and prioritised based on the cybersecurity expert's propositions (third author). When a user selects one coach, s(he) has access to the self-assessment questions and relevant capability training content. Providing the content was based on the research into the training material [26], technical reports provided by Symantec and Ponemon, and meeting with experts. Technical reports provide updated cybersecurity solutions and statistics. At the end of each coach, users see summary information and are redirected to the dashboard. In the dashboard, they see the progress information, achieved scores, and new recommendations for cybersecurity practices and selection of the next coach(es).

Table 2. Operationalisation of SDT Constructs

| SDT Construct | CYSEC Function | Operationalisation |
|---|---|---|
| Relatedness | Dashboard: recommendations | Self-adaptation of recommendations to SMB profile and improvement progress. |
| | Dashboard: progress summary | Continuous feedback about progress and motivation. |
| | Work area: steps | Self-adaptation of recommended next improvements |
| | Offline | Personal workshops with SMBs for reflecting about improvement experience. |
| Belonging and Connectedness | Work area: action cockpit | Fostering of personal communication between CISO and employees. |
| | Offline | Personal workshops with SMBs for reflecting about improvement experience. |
| Rewards, threats, and deadlines | Dashboard: progress summary | Feedback about the defence strength built and knowledge acquired in the company, and persistence in working on cybersecurity ("fitness"). |
| | Work area: expert knowledge | Information about the importance of improvements, e.g. by referring to cyber risks that should be mitigated. |
| | Work area: action cockpit | Setting of calendar entries and emailing reminders to employees. |
| Knowledge | Dashboard: access to capability areas | Access to knowledge and recommendations for building cybersecurity in the SMB. |
| | Work area: expert knowledge | Presentation of knowledge and recommendations for building cybersecurity in the SMB. |
| Choice | Dashboard: recommendations | Presentation of the three top recommendations, offering choice about the next important improvements. |
| | Dashboard: access to capability areas | Presentation of capability areas, offering choice about the type of cybersecurity to build. |
| | Work area: action cockpit | Choice of deferring improvements with a calendar entry or bookmark and of involving employees by email. |

## 4 Study Design

The study aimed at evaluating whether CYSEC is useful and effective as a method of communicating cybersecurity expertise for enabling DIY cybersecurity assessment and capability improvement for SMBs. To achieve this aim, we designed a deductive multi-case study and used observation as the main method for data collection [27]. Case studies are common in information systems research and cybersecurity [20].

For planning the case study, a study protocol was developed and sent to the participating SMBs. Before conducting the case studies, a pilot workshop was performed for a start-up project that involved the second and third researchers and a developer. The pilot allowed to identify and resolve initial problems in the study design. The selection of the cases was based on the availability of the SMBs. It has been done in two steps. At first, data collection was based on four SMBs, and during the study (project lifetime), two more SMBs were included. Based on Yin [27], when using a multiple-case design, the number of case replications is essential instead of sampling logic, and the model of generalisation is analytic generalisation when we have a developed theory as a template. The selected SMBs have security resources, working in the software industry, and their CISOs have a level of expertise in security. The CISOs' behaviour was the unit of analysis. Table 3 presents our SMBs' demographics. Based on the EU Commission definition, companies with < 50 employees and annual turnover ≤ €10 million are small, and those with < 250 employees and annual turnover ≤ €50 million medium.

### 4.1 Research Questions and Case Selection

The following two research questions were analysed. RQ1 related the communication method to the user's motivation and adoption of cybersecurity recommendations. RQ2 reflects users' evaluations of the method and tool after the actual usage.

RQ1: *How do the CYSEC dashboard and work area features influence the effectiveness of communicating cybersecurity to motivate users' adoption of desired behaviour?* To improve SMBs' cybersecurity capability, they need to adopt good cybersecurity practices. Since CYSEC provides security experts' recommendations and training content, we are studying SMBs' security adoption through CYSEC communication. The factors of the study are based ont he SDT constructs and CYSEC features in Table 2. We evaluate the effectiveness of communication by observing users' behaviour, attention, comprehension, and theoretical cause-effect relationships [27]. RQ2: *Do the SMB human end-users perceive CYSEC to be acceptable and useful as a tool assisting DIY cybersecurity assessment and improvement?* Tool acceptance and perceived usefulness are significant for us since a problem for information security is the users' resistance to accepting security tools [17]. Therefore, in RQ2, we want to have the users' attitude.

**Table 3.** SMBs Demographics

| ID | Step | Size | Offices | Maturity (Some controls implemented) | Structure |
|---|---|---|---|---|---|
| 1 | 1 | Medium | 3 | Access control., network controls, backup, encryption | CEO, security team, employees |
| 2 | 1 | Medium | 3 | Password mgmt., patch mgmt., encryption, (training) security best practices for developers | CEO, security team, employees |
| 3 | 1 | Small | 2 | Password mgmt., Code inspection | CEO, security team, employees |
| 4 | 1 | Small | 1 | Access control, patch mgmt., | Professors, manager, Security team, users |
| 5 | 2 | Small | 1 | Using a firewall, Access control | CEO, manager, employees |
| 6 | 2 | Small | 1 | Password mgmt., | CEO, managers, developers |

### 4.2 Workshop Design and Meeting with Companies

Each workshop started with an explanation of the study, steps, and relevant objectives. To collecting honest responses [28], the researcher emphasised that the collected data would be applied anonymously for academic purposes and then obtained the subjects' consent. During the workshops, the researcher took notes about how the subjects interacted with the tool and asked them to "think aloud" [29] and explain their expectation.

The workshops had four parts and four tasks for the CISO in the SMB. In part 1, the CISOs characterised their companies. In part 2, and to understand the main user's capabilities, the CISOs answered three questions about their level of education, experience in cybersecurity, and their roles. The responses to the questions were used to confirm the suitability of the selected case for the study. In part 3, the user applied the CYSEC. In part 4, the observation was supplemented with a post-observation quantitative questionnaire (Table 4). The result section presents the process of the workshops.

## 5    Results

In this section, we present the process of data collection in the 1-day workshops.

Company 1: two people (CISO) participated in the workshop, which took 26 minutes. This workshop took a short time since the users only provided their feedback at the end of the workshop. The researcher had the lowest degree of interaction in this workshop. It seems they were familiar with the topics and questions, so they refer to the training content only two or three times.

Company 2: Two people from a team of cybersecurity participated. The workshop took one hour and forty-three minutes. The users referred to the training content several times when they were not able to understand the actual goal of the questions. The users stated that the training content for some questions is not completely correct.

Company 3: Only one person (CISO) participated. The workshop took two hours and thirty-five minutes. The user explained that their SMB had not managed any cybersecurity training courses and security awareness issues are usually sent to the employees by email. During the workshop, the user experienced several system time-outs. So, the user required to log in several times. This issue was due to a bug in the tool, and it distracted the user's attention to some extent. So, the researcher needed to interact with the user. For this SMB, confidentiality issues were very critical.

Company 4: A team of seven IT specialist participated. The workshop took two hours and a half. The users went through the coaches the same as the first workshop. The distinctive feature of this workshop was the discussion about each question between participants to find suitable answers based on the company's requirements. The users applied training content too much to understand the questions. For this SMB dynamic, and reliable training material was significant.

Company 5: Only one person (CISO) participated. The workshop took thirty-four minutes. The user stated that their SMB has no chain of management. In this workshop, the user emphasised that some of the questions have no suitable options for the answer.

Company 6: Two people from a team of cybersecurity participated. The workshop took forty minutes. People in the workshop had a language problem, and they used Google translate to understand the content. So, the researcher needed to interact with the users. The users went through some of the coaches.

In all workshops, the mechanism of the tool usage was almost the same and the users referred to the training content mostly when they had a problem to understand questions. The meaning of some of the questions was not clear for the CISOs. Also, all of them required a summary and recommendation after finishing each coach.

## 6 Analysis

For answering the research questions, we explain the researcher's observations of the tool usage based on SDT- specific features of the CYSEC, which indicated in Table 2.

### 6.1 Factors Influencing Cybersecurity Communication (RQ1)

According to the workshop results, the CYSEC dashboard and work area features affected the users' motivation for the adoption of desired behaviour. These features together can facilitate security management and inter-organisational connection between the CISO and the employees and support self-efficacy and capability improvement in the SMBs. The participating SMBs learned about cybersecurity and adopted practices and controls when the immediate perceived learning experience was good.

**Available Expert Knowledge (CYSEC Work Area).** As suggested by SDT, we have observed that improving self-efficacy had a positive impact on the users' self-motivation. The immediate relevance of the training material provided by the CYSEC tool affected the subjects' decision to study the training material. Still, even though the participants did not have any systematic cybersecurity course so far, none of them read the full knowledge texts provided by CYSEC. Instead, the subjects studied training material when they did not understand a question or wanted to know more about a topic to select suitable answers. *Company 4: "We need practical instructions and steps to help us solve our issues and not general ideas." Company 2: "some materials are not relevant to the right topics/questions."* When responding to some of the questions, they referred to the training content occasionally and only to find a specific issue. For instance, to explore how they can measure the strength of password based on a tool.

Perceived reliability, expert support, clarity, and local language support of the content were important in the sense that the lack of these quality attributes hindered participants from accepting training input. *Company 4: "Parts of the content are not clear enough for us, and some materials (statistics) are not reliable since they are not covering many security experts and SMBs' opinions." Company 2: "Training content needs to present the severity of threats clearly." Company 2: "Since the coaches are in English, it might be possible for some SMBs to be unwilling to apply the tool." Company 6: "Can you please tell me what actually you mean* [even after using Google translate for the training content].*"* The quotes here can demonstrate the users' awareness and attention to the training content with respect to the effectiveness of the communication.

**Assessment Questions for Next Step Improvement (CYSEC Work Area).** As suggested by SDT, choice supported autonomy and self-motivation. However, the options should always be relevant and adapted to the IT infrastructure and operations of the company. The participants looked for questions that fit their interest and perceived needs. If they could not find such questions, they explained their needs by referring to their company's characteristics (assets). Moreover, they wanted the coach to adapt the questions to the SMB's characteristics specified with preceding answers. *Company 3: "here, the tool should provide a lot of questions to cover different operating systems."*

Also, the users wanted each question to be answerable with options for the response that explained the situation precisely. The researcher observed that the users did not answer some questions or selected an imprecise option due to lack of suitable choices.

*Company 2 about the question* "Do your users use any other authentication method, such as fingerprint or 2-factor authentication, to control access to your sensitive systems:" "*for some systems yes, for some systems and users no. I cannot answer properly, so I select the answer* [Yes, for some systems]*." Company 5: "I need an option between yes and no."* The quality of the questions also influenced the efficiency of the users. Some of the questions were perceived to be confusing. When being confronted with such questions, the participants looked for training content, searched the Internet for the topic, or asked the researcher for clarification. *Company 4: "some of the questions are confusing and not clear enough for us." Company 2: "I could not understand if the question is related to servers or employees' computers."* The quotes here can demonstrate the users' comprehension and their ability to answer the assessment questions. Their behaviours and comments refer to the effectiveness of communication.

**Action Cockpit (CYSEC Work Area).** As suggested by SDT, we have observed that respect of the belonging within the organisation is important. While the CISOs answered most training and awareness questions, they communicated with their colleagues to find the answer for some of the questions. The CISO in Company 2 made a call to answer some questions. When he could not find his colleague, he highlighted the question for future consideration. Also, in Company 4, the IT specialist had a discussion to find the best answers for some questions based on their company requirements.

**Recommendations (CYSEC Dashboard).** As suggested by SDT, self-efficacy and relatedness had a positive impact on users' autonomous motivation. All SMBs wanted to receive feedback after finishing a coach and recommendations for next improvements. *Company 1: "we need a summary and recommendation after each coach."*

**Access to Capability Areas (CYSEC Dashboard).** As suggested by SDT, choices support users' autonomy and motivation. Still, a recommended order was appreciated by the SMBs, even though the order must not be enforced. Some of the participants selected capability areas based on reflected priorities or requirements. *Company 2 on which capability do you want to select to answer? "no preference [for the capabilities]. we can answer based on the list."* Other participants followed different orders, and one looked for a specific capability area that was not available in the list of capability areas.

**Progress Summary (CYSEC Dashboard).** None of the participants was intrinsically motivated; SDT's model of introjected regulation was best explaining the participants' adoption behaviour. The participants wanted feedback about their performance, and it was important that the feedback was credible. The gamification elements of showing progress impacted the users' motivation. All the participated CISOs in this study wanted to have a simple summary which indicated their company's progress.

### 6.2 Acceptance and Usefulness of CYSEC (RQ2)

The answer to RQ2 is based on the supplementary data collected at the end of each study about the users' attitudes. Users evaluated the usefulness of the tool by responding to a five-level Likert scale questions about the tool usefulness (low [L], rather low [RL], medium [M], rather high [RH], high [H]) and justified their evaluation. The follow-up questionnaire (Table 4) aimed at understanding users' attitudes about tool acceptance and usefulness. However, the short survey and small number of SMBs were not statistically significant for analysis. Company 1: "*CYSEC is easy to use and useful.*" Company 2: "*the severity of the threats should be more visible in the training content.*" Also, Company 2 stated "*CYSEC needed to be evaluated by our employees.*" Company 3: "*I have not referred too much to the training content, so I prefer not to evaluate the second question.*" Moreover, this company explained that "*the confidentiality issues and the lack of some relevant questions in the specific topics influence our evaluation.*" Company 4: "*although CYSEC is easy to use, the instructions in the training content are not easy to implement and practical for us.*" Company 5: "*CYSEC provides lots of training material in one place. However, the gamification elements of showing progress need to be more transparent.*" Company 6: "*the logic behind the questions needs to be improved to better support adaptation.*" CYSEC usefulness was perceived to be high by Company 1, Company 4 and Company 6, rather high by Company 2 and Company 5, and medium by Company 3. The lowest rank (medium [M]) chosen by Company 3 put a strong emphasis on a) confidentiality before and during the study and b) the lack of relevant questions in specific advanced topics. This result indicates that a tool like CYSEC that is based on self-assessment and tailored training modules was accepted as a do-it-yourself approach allowing SMBs to manage most capabilities.

**Table 4.** Questionnaire for the workshops' part 4.

| **Part 4 Questionnaire** | ID 1 | ID 2 | ID 3 | ID 4 | ID5 | ID6 |
|---|---|---|---|---|---|---|
| Have you been aware of these threats/vulnerabilities? (Content) | RH | RH | H | RH | RH | H |
| How do you evaluate the quality of the information in the training content? | RH | RH | - | RH | RH | H |
| Does the training content send a clear message about the threat severity or your vulnerability? | RH | M | M | RH | RH | RH |
| Are the instructions of the training content doable? | H | RH | M | M | H | H |
| How easy is applying CYSEC (easy to use)? | RH | - | M | H | H | H |
| How useful is applying CYSEC to improve your security awareness and capability? | H | RH | M | H | RH | H |

## 7 Discussion

We have presented an approach for allowing SMBs to improve their cybersecurity in a DIY fashion. This result allows moving from a bespoke, consultancy-centred advisory to an automated model of communicating cybersecurity knowledge that is scalable yet individualised, hence allows to serve many SMBs with little effort. Our approach implements the self-determination theory [21], which describes the motivation for achieving outcomes and the factors influencing such motivation. SDT is not the only relevant theory, however. Protection motivation theory [19] is one of the widely used theories and would offer an alternative for method and tool design, allowing us to focus on motivating the users to protect their assets and company. We have chosen to implement SDT first as, in our understanding, the improvement of cybersecurity also concerns the creation of new capabilities in the organisation: installing and configuring tools, establishing policies, and training employees beyond just protecting an asset.

In a multi-case study, we offered insights on the actual use of our designed method and tool in real-world SMB settings. Such validation goes beyond just evaluating intentions as in [19] or theoretical relationship as in the common survey-based studies. The presented work is the first step towards operationalising SDT and using it to achieve an impact on practice. As a result, we have discovered issues for future research that would not have been discovered otherwise. The findings of the study suggest that one challenge of motivating and supporting SMBs is the choice of knowledge that is being communicated. The wrong knowledge, extraneous security awareness details, or knowledge gaps reduce motivation and influence the adoption of security recommendations. Also, we discovered a potential barrier that should be addressed by future research: confidentiality. While SDT emphasises relatedness, we observed resistance to documenting and sharing security-related information both within and among companies. Alleviating confidentiality worries is crucial for improving the method's success.

Following Yin [27], our study has the following threats to validity. **Construct validity:** are the operational measures for the concepts being studied correct? Our choice of constructs is based on SDT (Table 1) that we implemented in the CYSEC tool (Table 2) and discussed with the study participants. We described how we had implemented the constructs and offered a chain of evidence between the answers to the research questions and the data collected in the workshops. To ensure that the results reflect not only our subjective impressions, we did member checking.

**Internal validity:** can the cause-effect relationships in SDT be distinguished from bogus relationships? We used pattern matching and explanation-building for addressing internal validity threats. We identified relevant observations and feedback collected in the workshops and evaluated which ones spoke for a relationship, respectively against, resulting in the reported analysis. We also asked for what could have influenced the assessment, for example, whether the participants had cybersecurity training before the workshop to rule out the influence of such expertise. A longitudinal study, e.g., based on follow-up surveys, could be a research strategy for evaluating whether and how the SMBs change their practices with extended use of the CYSEC tool.

**External validity:** can the study be generalised? The participating SMBs were diverse but still had similarities: all were active in the digitisation, had a CISO without deep expertise in cybersecurity but several years of experience, and did not provide any training to their employees. Since they had some knowledge in cybersecurity, they were able to answer the questions without continually referring to the training content and succeeded to implement some security controls. However, this study has not covered a) SMBs with considerable expertise in the cybersecurity b) SMBs that hardly use IT, and c) SMBs without any budget and personnel for cybersecurity. The study results may change for such SMBs because of different knowledge needs and relevance of assessment questions and improvement recommendations.

**Reliability**: can the study be replicated with the same results? We developed and piloted a case study protocol that we applied in the main study. The protocol ensured that the operational steps were clear and that each SMB had enough time to use the tool without distraction. All the steps were traced in a case study database. To strengthen the chain of evidence, we presented our findings to the participating SMBs' subjects and cybersecurity experts in a formal meeting for correction (member checking).

## 8  Summary and Conclusions

The paper has evaluated the actual usage of the CYSEC, a do-it-yourself (DIY) security assessment and improvement method for small and medium-sized businesses (SMBs), through an explanatory multi-case study. This study followed a deductive approach and tested constructs drawn on the Self-Determination Theory to evaluate the impact of the method on the effectiveness of cybersecurity communication to the SMBs. We applied observation and feedback questionnaires for data collection.

The results support the influence of the following features on communication effectiveness and users' motivation: expert knowledge, self-adapting assessment question, action cockpit for connectedness in SMBs, self-adapting recommendations, provided capability areas, and progress summary. They empower SMBs in adopting and adhering to cybersecurity. The content, including questionnaires and recommendations, need to be presented in an easy-to-understand manner to improve users' competency. The assessment questionnaires and recommendations need to adapt to each specific SMB to increase autonomy. Also, for users' acceptance and adherence, CYSEC needs to consider the company confidentiality seriously. Users are emotionally connected to the SMBs' data do not want to share their information, especially about vulnerabilities, with the tool or third parties outside the SMB. Thus, confidentiality, trust, and relatedness may influence security communication and tool acceptance positively.

## Acknowledgements

This work was made possible with funding from the European Union's Horizon 2020 research and innovation programme under grant agreement No 740787 (SMESEC) and the Swiss State Secretariat for Education, Research and Innovation (SERI) under contract number 17.00067. The opinions expressed and arguments employed herein do not necessarily reflect the official views of these funding bodies.

## References

1. Muller, P., Julius, J., Herr, D., Koch, L., Peycheva, V., McKiernan, S., Hope, K.: Annual report on European SMEs 2016/2017: Focus on self-employment. European Commission (2017).
2. Caldwell, T.: Securing small businesses – the weakest link in a supply chain? Computer Fraud & Security, 5–10 (2015).
3. Ntouskas, T., Papanikas, D., Polemi, N.: A Collaborative System Offering Security Management Services for SMEs/mEs. In: Georgiadis, C.K., Jahankhani, H., Pimenidis, E., Bashroush, R., and Al-Nemrat, A. (eds.) Global Security, Safety and Sustainability & e-Democracy. pp. 220–228. Springer, Berlin Heidelberg (2012).
4. Goucher, W.: Do SMEs have the right attitude to security? Computer Fraud & Security, 18–20 (2011).
5. Gupta, A., Hammond, R.: Information systems security issues and decisions for small businesses: An empirical examination. Info Mngmnt & Comp Security. 13, 297–310 (2005).
6. Mijnhardt, F., Baars, T., Spruit, M.: Organizational characteristics influencing SME information security maturity. Journal of Computer Information Systems 56(11), 106-115 (2016).

7. Kurpjuhn, T.: The SME security challenge. Computer Fraud & Security, 5–7 (2015).
8. Valli, C., Martinus, I.C., Johnstone, M.N.: Small to Medium Enterprise Cyber Security Awareness: an initial survey of Western Australian Business (2014).
9. Brunner, M., Sillaber, C., Breu, R.: Towards Automation in Information Security Management Systems. In: 2017 IEEE International Conference on Software Quality, Reliability and Security (QRS), pp. 160–167. IEEE, Prague, Czech Republic (2017).
10. Furnell, S.M., Gennatou, M., Dowland, P.S.: A prototype tool for information security awareness and training. Logistics Information Mngt. 15, 352–357 (2002).
11. Dhillon, G., Torkzadeh, G.: Value-focused assessment of information system security in organizations. Information Systems Journal. 16, 293–314 (2006).
12. Cranor, L.F.:A framework for reasoning about the human in the loop. In: UPSEC'08 Proceedings of the 1st conference on usability, psychology, security, Berkeley, USA (2008).
13. Pahnila, S., Siponen, M., Mahmood, A.: Employees' Behavior towards IS Security Policy Compliance. In: 40th Annual Hawaii International Conference on System Sciences (HICSS'07), pp. 156–166. IEEE, Hawaii, USA (2007).
14. Rhee, H.-S., Kim, C., Ryu, Y.U.: Self-efficacy in information security: Its influence on end users' information security practice behavior. Computers & Security. 28, 816–826 (2009).
15. Shojaifar, A., Fricker, S.A., Gwerder, M.: Elicitation of SME Requirements for Cybersecurity Solutions by Studying Adherence to Recommendations. REFSQ Workshops (2018).
16. Albrechtsen, E., Hovden, J.: The information security digital divide between information security managers and users. Computers & Security, 28, 476–490 (2009).
17. West, R.: The psychology of security. Communications of the ACM 51(4), 34-40 (2008).
18. Hayes, J.: The theory and practice of change management. Palgrave, New York (2002).
19. Menard, P., Bott, G.J., Crossler, R.E.: User Motivations in Protecting Information Security: Protection Motivation Theory Versus Self-Determination Theory. Journal of Management Information Systems 34(4), 1203–1230 (2017).
20. Pham, H.C., Pham, D.D., Brennan, L., Richardson, J.: Information Security and People: A Conundrum for Compliance. AJIS 21, 1-16 (2017).
21. Ryan, R.M., Deci, E.L.: Self-Determination Theory and the Facilitation of Intrinsic Motivation, Social Development, and Well-Being. American Psychologist. 55, 68-78 (2000).
22. Deci, E.L., Ryan, R.M.: Self-determination theory: A macrotheory of human motivation, development, and health. Canadian Psychology 49(3), 182–185 (2008).
23. Deci, E.L., Ryan, R.M.: The General Causality Orientations Scale: Self-determination in personality. Journal of research in personality 19(2), 109-134 (1985).
24. Deming, W.E.: Elementary principles of the statistical control of quality. Nippon Kagaku Gigutsu Remmei: Japanese Union of Science and Engineering (JUSE) (1951).
25. Ozkan, B.Y., Spruit, M.: A questionnaire model for cybersecurity maturity assessment of critical infrastructures. In: Fournaris, A.P., Lampropoulos, K., and Marín Tordera, E. (eds.) Information and Operational Technology Security Systems. pp. 49–60. Springer International Publishing, Cham (2019).
26. Gardner, B., Thomas, V.: Building an Information Security Awareness Program: Defending Against Social Engineering and Technical Threats. Elsevier/Syngress, Amsterdam (2014).
27. Yin, R.K.:Case study research:design and methods.4th edn.Thousand Oaks,CA: Sage (2009).
28. Bennett, R.J., Robinson, S.L.: Development of a measure of workplace deviance. Journal of Applied Psychology. 85(3), 349–360 (2000).
29. Runeson, P., Höst, M., Rainer, A., Regnell, B.: Case Study Research in Software Engineering: Guidelines and Examples. John Wiley & Sons, Inc., Hoboken, USA (2012).